\documentclass[a4paper]{article}

\usepackage{INTERSPEECH2020}
\usepackage{bm}
\usepackage{multirow}

\title{Alzheimer's Dementia Detection from Audio and Text Modalities}
\name{Edward L. Campbell$^1$,  Laura Doc{\'i}o-Fern{\'a}ndez$^1$, Javier Jim{\'e}nez Raboso$^2$, Carmen Garc{\'i}a-Mateo$^1$}
\address{
  $^1$GTM research group,  
  AtlanTTic Research Center, 
  University of Vigo, Spain
\\
  $^2$acceXible
}
\email{ecampbell@gts.uvigo.es, ldocio@gts.uvigo.es,  javij@accexible.com, carmen.garcia@uvigo.es}

\begin{document}

\maketitle
\begin{abstract}

Automatic detection of Alzheimer's dementia by speech processing is enhanced when features of both the acoustic waveform and the content are extracted. Audio and text transcription have been widely used in health-related tasks, as spectral and prosodic speech features, as well as semantic and linguistic content, convey information about various diseases. Hence, this paper describes the joint work of the GTM-UVIGO research group and acceXible startup to the ADDReSS challenge at INTERSPEECH 2020. The submitted systems aim to detect patterns of Alzheimer's disease from both the patient’s voice and message transcription. Six different systems have been built and compared: four of them are speech-based and the other two systems are text-based. The x-vector, i-vector, and statistical speech-based functionals features are evaluated. As a lower speaking fluency is a common pattern in patients with Alzheimer's disease, rhythmic features are also proposed. For transcription analysis, two systems are proposed: one uses GloVe word embedding features and the other uses several features extracted by language modelling. Several intra-modality and inter-modality score fusion strategies are investigated. The performance of single modality and multimodal systems are presented. The achieved results are promising, outperforming the results achieved by the ADDReSS's baseline systems.\\
     
\end{abstract}


\section{Introduction}

The Alzheimer's disease (AD) is a neurodegenerative illness that represents the most common cause of dementia in the world. It is provoked by the damage of neurons involved in thinking, learning and memory. This disease has three main stages. In the first one, called preclinical, patients do not present clear symptoms because the brain initially compensate for them, enabling individuals to continue to function normally. The second one is defined as Mild Cognitive Impairment (MCI). In this stage, patients show greater cognitive decline then expected for their age, having problems to express and connect ideas. However, these changes may be only noticeable to family members and friends. The critical phase is the last one, which is called as dementia. It is characterized by noticeable memory, thinking and behavioural symptoms that impair a person's ability to function in daily life \cite{alzheimer20192019}\cite{nestor2004advances}.

Common signs of AD are related to problems with uttering words; consequently, people with AD may have trouble following or joining a conversation, they may stop in the middle of a sentence and have no idea how to continue. As a result, analysis of speech and its transcription may represent a suitable mechanism for detecting the AD during the second or third stage of the disease \cite{alzheimer20192019} \cite{ahmed2013connected}.
According to the literature\cite{meilan2014speech}\cite{lopez2015automatic}\cite{fraser2016linguistic}, using information from the patient's voice, as well as from its transcription would ease the early AD detection task.

In this work, four speech-based systems and two text-based systems are proposed for automatic distinction between patient's with and without AD.
Those based on the speech signal are compound by four approaches to represent their spectral and prosodic content, namely: i-vector and x-vector embeddings, rhythmic features, and statistical functionals. The first three use super vector machine (SVM) as classifiers and the last one uses a linear discriminant analysis (LDA) classifier. As for systems that use speech transcriptions, one relies on GloVe word embedding features \cite{pennington2014glove} using a recurrent neural network (RNN) as classifier, and the other one is based on features extracted by language modelling, using a SVM as classifier. Finally, an intra-modality and inter-modality score fusion strategy was done to improve the final results. 

All the individual systems, and its fusion, are evaluated on the AD recognition task within the framework of the ADReSS Challenge \cite{bib:LuzHaiderEtAl20ADReSS}. This challenge targets the AD detection  using spontaneous speech. The data used in the Challenge consists of speech recordings, and their transcripts, corresponding to the description of the Cookie Theft picture. Specifically, it is a selection of Alzheimer and control patients from of the DementiaBank’s Pitt corpus \footnote{http://dementia.talkbank.org/}.

The rest of the paper is organized as follows. Sections \ref{sec:speech-systems} and  \ref{sec:text-systems} describe the speech-based and text-based systems, respectively. Section \ref{sec:experimental_framework} outlines the experimental framework. The experimental results are exposed and discussed in Section 
\ref{sec:results}. Finally, Section \ref{sec:conclusions} draws some conclusions and future work.

\section{Speech-based systems}
\label{sec:speech-systems}
In this section, a description of speech feature extraction strategies and classification methods is done, with special attention to different statistical features extracted from the patient's voice.

\subsection{Speech embedding features}

Speech embedding features are considered as state-of-art speech representation for speaker recognition application. These speech representations can be applied for AD detection, as long as they preserve those spectral patterns in the speaker's voice that allow the distinction between patient with and without AD. In this paper, two strategies were analyzed, the first one uses the i-vector paradigm \cite{dehak10}, and the second one uses an x-vector \cite{PoveyXVectors} based representation. The main characteristics of these approaches are briefly described below.

\begin{list}{\bf{A.}}{\leftmargin=0em \itemindent=1.5em}
  \item {\bf{\emph{i-vectors}}}

To extract the i-vectors an universal background model (UBM) and a total variability matrix $\mathbf{T}$ model must be trained. 
As speech parameterization these models use 13 perceptual linear prediction (PLP) cepstral coefficients, combined with two pitch-related features (F0 and voicing probability) \cite{ghahremani2014}. This features are augmented with their delta and acceleration coefficients, leading to vectors of dimension 45.
These features were chosen since that combination achieves a representation of speech that includes spectral information and prosodic features such as rhythm or intonation that are embedded in the fundamental frequency. 
The UBM was a diagonal covariance, 512-component gaussian mixture model (GMM), and was trained with data from outside this task; and $\mathbf{T}$ was trained using the task training data. The dimension of the i-vectors was set to 125 and they were length normalized.

\end{list}

\begin{list}{\bf{B.}}{\leftmargin=0em \itemindent=1.5em}
  \item {\bf{\emph{x-vectors}}}

A pretrained time delay deep neural network (TDNN) \footnote{http://kaldi-asr.org/models.html}, with 5 time delay layers and two dense layers, and trained to discriminate between speakers, was used. The network was implemented using the nnet3 neural network library in the Kaldi Speech Recognition Toolkit \cite{PoveyKaldi} and trained on augmented VoxCeleb 1 and VoxCeleb 2 datasets \cite{NagraniCXZ20}. The input to the TDNN are 30 mel-frequency cepstral coefficients (MFCC), and the embeddings are extracted from the first dense layer with a dimensionality of 512. The output of this layer (x-vector) is first projected using latent discriminant analysis (LDA) into a 200 dimensional space and then length normalized.

\end{list}

Instead of extracting an i-vector or x-vector (embeddings) to represent the entire audio signal, a set of these vectors is obtained applying a sliding window. In this way, each audio file is represented by a certain number of embeddings, which are then used for classification.

Both systems use as classifier an SVM with a linear kernel. Since a number of embeddings are extracted from each audio, there will also be a set of classification results (one for each embedding), which must be combined to obtain a patient's classification in AD or non-AD. In this work, the mean of the classifications was used as score for the final decision.

\subsection{Statistical functionals}

This system represents the audio waveform with a set of vectors called functionals. These are mainly statistical and regression features extracted from a wide variety of low-level acoustic attributes using the openSMILE \cite{EybenWS10} software. Specifically, the set of functionals proposed in the AVEC 2013 evaluation (Audiovisual Emotion Recognition Challenge and Depression) \cite{avec2013} was used, which consist of 2,268 descriptors extracted from 32 low-level features of the audio signal related to energy, the spectrum, volume and tone. 
 
As the number of descriptors is very high in relation to the amount of data available, a selection of those most relevant was done. For this, a correlation based feature subset selection (CFS) algorithm \cite{Hall1998} was used, resulting in 57 functionals related with the low level descriptors shown in Table \ref{tab:57functionals}.
The functionals were extracted using a sliding window of 3 s and an overlap of 1 s.
As with the i-vectors and x-vectors this system uses an SVM with a linear kernel as classifier.

\begin{table}[ht]
\centering
\caption{Low level descriptors in the set of 57 functionals}
\vspace{1mm}
\begin{tabular}{l} \hline
\textbf{Energy and spectral features}\\ \hline
Loudness, MFCC and delta-MFCC, \\
energy in bands 250--650Hz and 1 kHz--4 kHz,\\
25\% spectral roll-of points, \\
spectral flux, entropy, skewness,\\
psychoacousitc sharpness, harmonicity, flatness\\
\hline
\textbf{Voicing related}\\ \hline
$F_0$,voicing probability, jitter \\
\hline
\end{tabular}
\label{tab:57functionals}
\end{table}

\subsection{Speech fluency}

A common pattern in patient with AD is the lack of speaking fluency, being the rhythm a viable clue to detect that behavior. However,  prosodic information (e.g., mean energy) was also used because the correct pronunciation of words does not only depends on the rhythm but intonation, tone and stress. For this system, the selected parameters were based on \cite{meilan2014speech} \cite{ajili2018impact}, and they are as follows: 

\begin{itemize}
\item Number of syllables
\item Rate of speech (syllables / original duration)
\item Speaking duration
\item Average fundamental frequency
\item Median of fundamental frequency
\item Minimum fundamental frequency 
\item Pronunciation posterior probability

\item Average voice interval duration
\item Average duration of pairs \footnote{Pairs: consecutive voiced and unvoiced segments}
\item Mean energy
\item The ratio between the energy mean and its standard-deviation

\end{itemize}

The extraction process was done using  the Python library My-Voice Analysis \footnote{https://github.com/Shahabks/my-voice-analysis}, the voice activity detection of the SIDEKIT software \cite{larcher2016extensible}, and our own algorithms.  

The classification process was done by the LDA algorithm, projecting the rhythmic feature vector into an one-dimensional space where every projected point represents a classification score.

\section{Text-based systems}
\label{sec:text-systems}
Two different text-based approaches were analysed. The first one uses the manual transcripts to extract linguistic information for creating the input features of a classifier. The second one is a sequential model based on deep learning, which classifies directly from the sequence of GloVe word embeddings.

\subsection{Text preprocessing}
The transcripts contained in the dataset are in CHAT format \cite{macwhinney2000}, which facilitates speech annotation and analysis. They include the transcript of both the subject and the investigator in charge of the test, as well as additional non-speech annotations: times, researcher interventions, errors, morphological or syntactic analysis.

Transcripts are divided into several interventions, i.e. sentences or parts of complete sentences with meaning, and this granularity has been maintained in the preprocessing. Metadata, researcher interventions and linguistic analysis included in the files have been removed. We keep only subject’s words as input for the classifiers, as it better reflects the real-world task of detecting Alzheimer’s based only on speech analysis.

\subsection{Sequential model}
The first approach consists of training a RNN at intervention level, by classifying if a given sentence belongs to an AD subject. Then, the probability that the subject had AD given all his/her interventions is computed as the mean of the probabilities of each intervention. By using the interventions as independent samples the training dataset is increased to a size of 1492 records from the 108 subjects. In the training process, words are tokenized, and interventions are padded up to 20 tokens.

The RNN is composed of 3 layers: an Embedding layer with pre-trained weights of GloVe 50-dimensional representation of words, a long short term memory (LSTM) layer with 4 units and an output layer with 1 unit and sigmoid activation. Additionally, a Dropout at 10\% rate is performed in the Embedding and LSTM layers for regularization and to prevent overfitting. The model is trained for 10 epochs, using a batch size of 16 and Adam optimization.

\subsection{Model based on linguistic features}

In this approach, several linguistic features and indicators are built from subject’s interventions and, using this feature vector as input, an SVM is trained. Unlike the previous method, the classification here is performed at subject level, by considering the full transcript of each participant. 

Previous works \cite{fraser2016} have shown that certain linguistic features are useful for detecting AD using Cookie Theft test. Here 13 features have been built, grouped into 4 categories:
\begin{itemize}
    \item Extension information such as the number of interventions, number of words per intervention and mean word length.
    \item Vocabulary richness, by measuring the number of unique words used by the subject.
    \item Presence of key informational concepts: kitchen, mother, stool, boy and girl.
    \item Frequency of verbs, nouns, adjectives and pronouns from POS-tagging.
\end{itemize}

Each feature is then rescaled with min-max normalization in range [0, 1] and an SVM with radial basis function (RBF) kernel and C = 1.0 is trained, whose output is the probability that the subject had AD given the 13-dimensional feature vector.

\section{Experimental framework}
\label{sec:experimental_framework}

The training dataset \cite{bib:LuzHaiderEtAl20ADReSS} consists of the recordings and manual transcripts of 108 subjects performing the test known as Cookie Theft, whose objective is to describe an image. Out of the 108 participants, 54 are patients diagnosed with Alzheimer's. Table \ref{tab:train} shows the training data distribution in detail.

\begin{table}[th]
  \caption{ADReSS training dataset}
  \label{tab:train}
  \centering
  \resizebox{5cm}{!}{
  \begin{tabular}{ l l  r  r r } 
    \toprule
    & \multicolumn{2}{c}{\textbf{AD}} & \multicolumn{2}{c}{\textbf{non-AD}}\\  
   Age interval&Male&Female&Male&Female\\
    \midrule
$[50,55)$&1&0&1&0\\
$[55,60)$&5&4&5&4\\
$[60,65)$&3&6&3&6\\
$[65,70)$&6&10&6&10\\
$[70,75)$&6&8&6&8\\
$[75,80)$&3&2&3&2\\
\midrule
Total&24&30&24&30\\
    \bottomrule
  \end{tabular}}
\end{table}

The evaluation metrics for the AD classification task are: $Accuracy=\frac{TN + TP}{N}$, Precision $\pi= \frac{TP}{TP+FP}$, Recall $\rho=\frac{TP}{TP+FN}$ and $F_{1}=2\frac{{\pi}* {\rho}}{{\pi} +{\rho}}$.
\\

where N is the number of patients, TP, FP and FN are the number of true positives, false
positives and false negatives, respectively.

All the systems have been trained using leave-one-subject-out (LOSO) cross-validation strategy for measuring the generalization error. Therefore, models use 107 subjects as training data and are validated on the held-out subject.

\section{Results}
\label{sec:results}
This section presents the results in the AD classification for both the  leave-one-subject-out (LOSO) and test settings of the ADReSS challenge.

\subsection{Results for LOSO setting}
Table \ref{tab:result1} illustrates the individual results achieved by the four speech-based systems. These results show that the x-vectors, functionals, and fluency based systems have similar performance regarding the accuracy, as well as Area Under Curve (AUC). The i-vector was the weakest model, being the only one with an accuracy under 70\%, although it is still above the challenge baseline results \cite{bib:LuzHaiderEtAl20ADReSS}.

\begin{table}[th]
  \caption{Results of the proposed speech-based systems}
  \label{tab:result1}
  \centering
  \resizebox{8cm}{!}{ \begin{tabular}{ ll l l   l l l} 
    \toprule
&class & Precision & Recall & F1 Score & Accuracy & AUC \\
    \midrule\midrule

\multirow{2}{*}{i-vector} & non-AD & 0.7058 & 0.6666 & 0.6857 & \multirow{2}{*}{0.6944} & \multirow{2}{*}{0.6806} \\
                          & AD & 0.6842 & 0.7222 & 0.7027  \\\midrule
\multirow{2}{*}{x-vector} & non-AD & 0.7400&0.6851 &0.7115 & \multirow{2}{*}{0.7222} & \multirow{2}{*}{0.7615} \\
                           & AD & 0.7068& 0.7592& 0.7321 \\\midrule
\multirow{2}{*}{functionals} & non-AD & 0.7454&0.7592 &0.7522 & \multirow{2}{*}{0.7500} & \multirow{2}{*}{0.7435} \\
                        & AD & 0.7547&0.7407 & 0.7476 \\\midrule
\multirow{2}{*}{fluency} & non-AD &0.7450 &0.7037 &0.7238 & \multirow{2}{*}{0.7314} & \multirow{2}{*}{0.7613}\\
                        & AD & 0.7192&0.7592 & 0.7387 \\

    \bottomrule
  \end{tabular}  }
\end{table}

The results for both text-based systems are shown in the table \ref{tab:result2}. Concerning the RNN model, the AUC in held-out set is 0.8563. For a selected threshold probability, we also obtain an accuracy of 0.7407, recall of 0.8704 and F1-score of 0.7705. The false negative rate is 13\%. For the linguistic model, the AUC in held-out set is 0.7510. For a selected threshold probability the accuracy is 0.6852, recall of 0.7037 and F1-score of 0.6909. The false negative rate is 30\%.

\begin{table}[th]
  \caption{Results of the proposed text-based systems}
  \label{tab:result2}
  \centering
    \resizebox{8cm}{!}{ \begin{tabular}{ lll l l l l} 
    \toprule
&class & Precision & Recall & F1 Score & Accuracy & AUC \\
    \midrule\midrule
\multirow{2}{*}{RNN model} & non-AD & 0.7424&0.9074 &0.8166 & \multirow{2}{*}{0.7962} & \multirow{2}{*}{0.8563} \\
                          & AD & 0.8809& 0.6851& 0.7708  \\\midrule
\multirow{2}{*}{Linguistic model} & non-AD & 0.7017&0.7407 &0.7207 & \multirow{2}{*}{0.7129} & \multirow{2}{*}{0.7510} \\
                           & AD & 0.7254&0.6851 & 0.7047 \\

    \bottomrule
  \end{tabular}}
\end{table}

\begin{figure}[th]
  \centering
  \includegraphics[width=\linewidth,height=5.5cm]{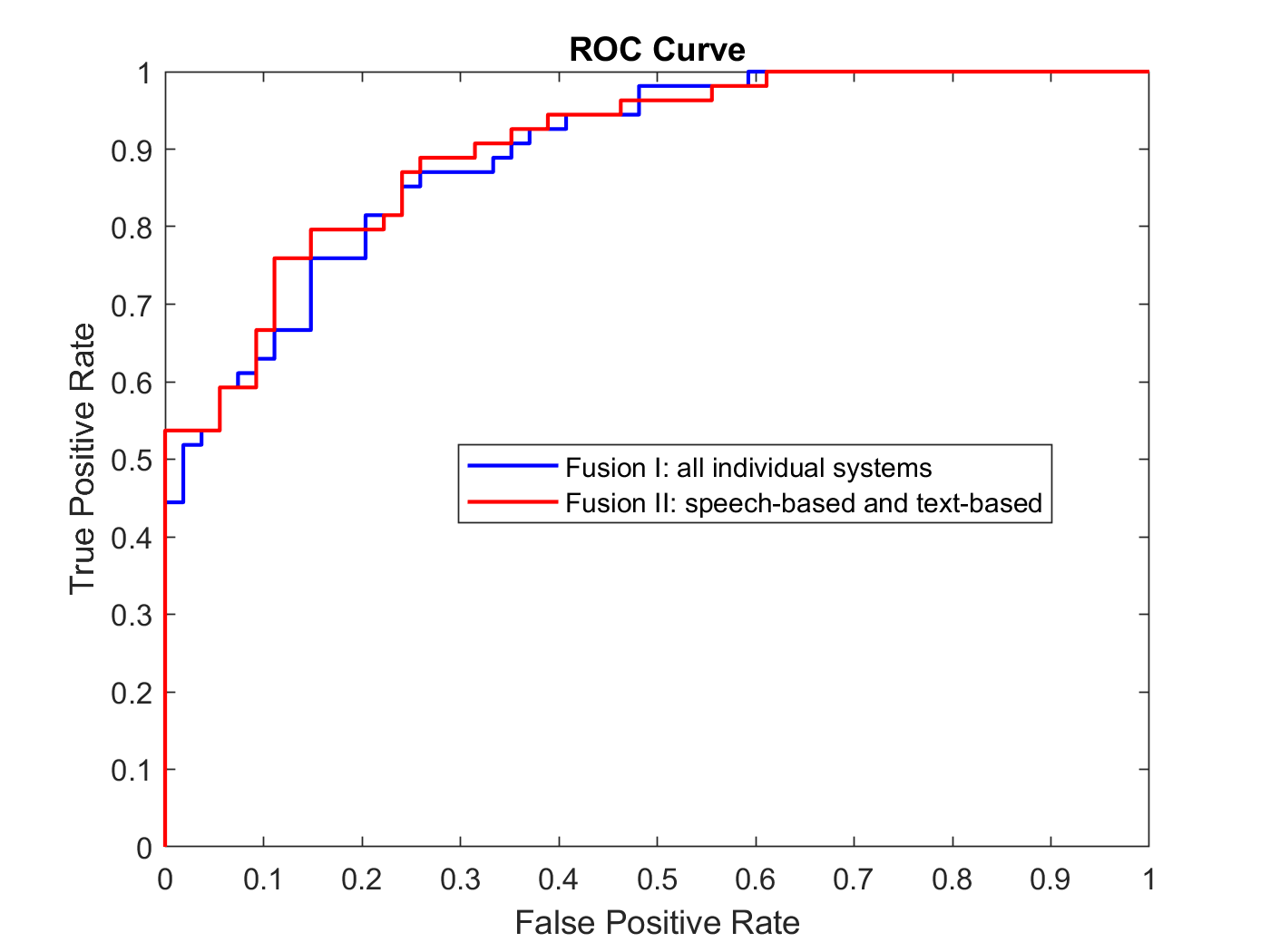}
  \caption{ROC curve of fusion}
  \label{fig:fusion-rocs}
\end{figure}

Moreover, two different score fusion strategies were carried out. In the first one (referred as Fusion I), the scores of every system was normalized by z-norm, and merged by the average. In the second one (referred as Fusion II), the same normalization strategy was used and the average was computed on the speech-based and text-based system scores separately. Finally, the new scores were again merged by the average. The figure \ref{fig:fusion-rocs} shows the ROC curves of the described fusion systems. The Fusion I model had an accuracy of 0.8056, a  F1-score of 0.8073, and an AUC of 0.8930. On the other hand, the Fusion II model presented an accuracy of 0.8241, a  F1-score of 0.8190 and an AUC of 0.9023. The results show that the fusion of both text-based and speech-based modality improves the detection of AD.

\begin{figure}[th]
  \centering
  \includegraphics[width=\linewidth,height=5.5cm]{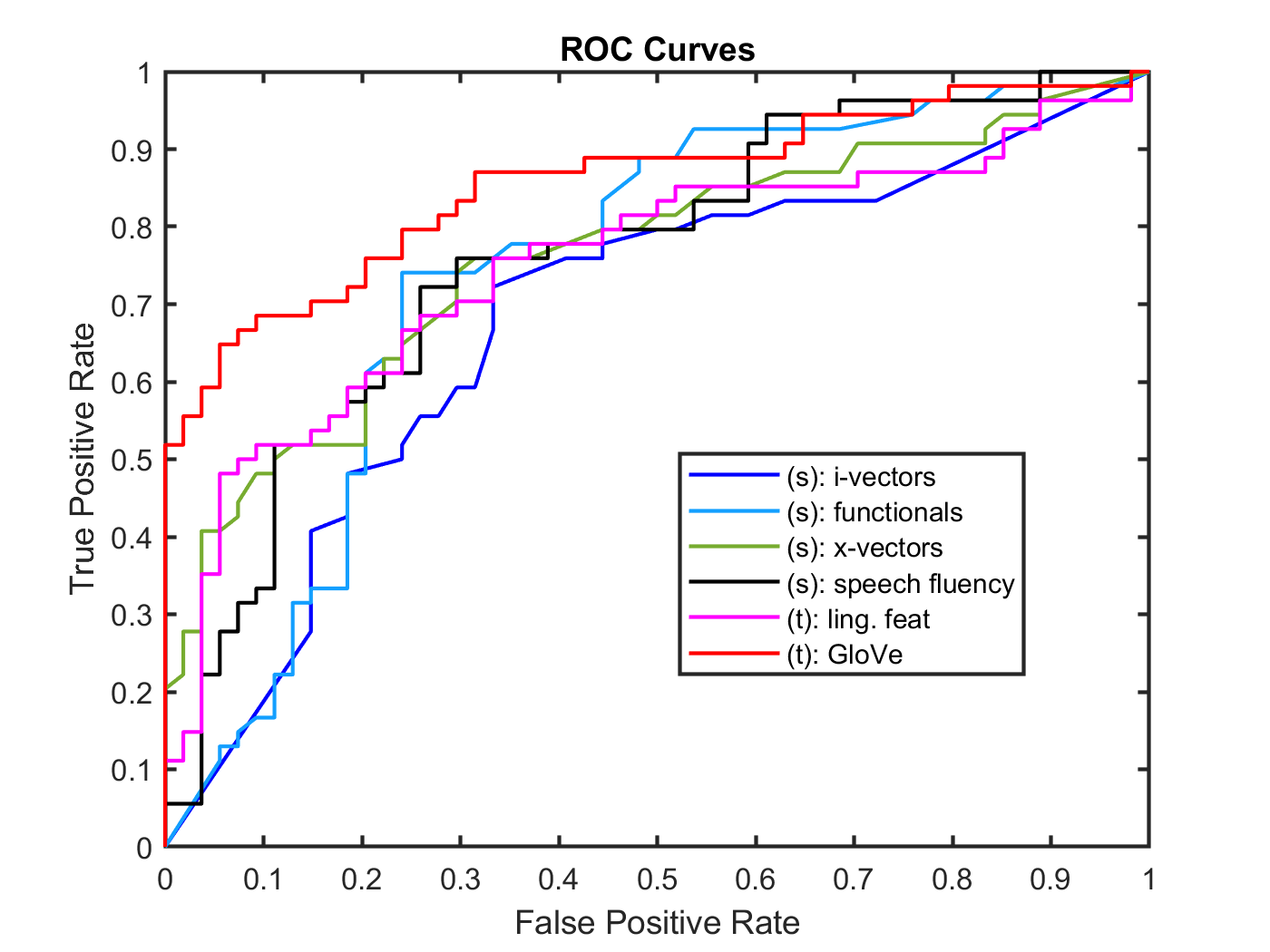}
  \caption{ROC curve of all systems}
  \label{fig:individual-rocs}
\end{figure}

For further insight, figures \ref{fig:individual-rocs} and \ref {fig:fusion-rocs} compares the ROC curves of all individual systems and their fusion, respectively.


\subsection{Results for test setting}
Based on the previous results, five systems were submitted to the challenge. The table \ref{tab:test} shows the performance achieved by them on the testing dataset, which consist of recordings of 48 subjects.

\begin{table}[th]
  \caption{Results on the testing dataset}
  \label{tab:test}
  \centering
  \resizebox{8cm}{!}{ \begin{tabular}{ ll l l   l l l} 
    \toprule
&class & Precision & Recall & F1 Score & Accuracy \\
    \midrule\midrule

\multirow{2}{*}{Fusion II} & non-AD & 0.8000 & 0.6667  & 0.7273 & \multirow{2}{*}{0.7500} \\
                          & AD & 0.7143 & 0.8333 &  0.7692 \\\midrule
\multirow{2}{*}{Fusion I} & non-AD & 0.8235&0.5833 &0.6829 & \multirow{2}{*}{0.7292} \\
                           & AD &0.6774 &0.8750 &0.7636  \\\midrule
\multirow{2}{*}{RNN model} & non-AD & 0.6667&1.0000 &0.8000 & \multirow{2}{*}{0.7500} \\
                        & AD & 1.0000&0.5000 &0.6667  \\\midrule
\multirow{2}{*}{fluency} & non-AD &0.6087 &0.5833 &0.5957 & \multirow{2}{*}{0.6042}\\
                        & AD & 0.6000&0.6250 &0.6122  \\\midrule
\multirow{2}{*}{x-vector} & non-AD &0.5553 &0.4167& 0.4762 & \multirow{2}{*}{0.5417}\\
                        & AD & 0.5333&0.6667 &0.5926  \\

    \bottomrule
  \end{tabular}  }
\end{table}

The results achieved by the RNN model were really similar in training and testing, which is a good sign of the lack of overfitting on the training stage as a result of a Dropout at 10\% rate performed in the Embedding and LSTM (Long Short Term Memory) layers. However, the submitted speech-based systems had a significant decrease in performance. This decline would be a consequence of a poor generalization capacity. As a result, the performance of the fusion systems also declines.

\section{Conclusions and Further work}
\label{sec:conclusions}
Two lines of work have been developed on the ADReSS Challenge dataset: one based on speech processing and the other based on text processing.

It is important to highlight the following points when assessing both solutions:

\begin{itemize}
    \item {\bf Performance:}    x-vector, functionals, and fluency speech features have shown a competitive performance in the LOSO cross-validation. However, their performance decreases on the test setting. This would be a result of a low generalization level achieved at the training stage. On the other hand, the text-based RNN model has obtained superior results both in the LOSO and testing settings, but with unbalanced values of recall and precision in the last one.  Nevertheless, the Fusion \textbf{II} strategy was a solution to that problem, presenting the same accuracy that the RNN model but with more balanced measures. As a result, multimodal systems demonstrate to be a better strategy for the detection of AD than individual sytems.

      \item {\bf Complexity:}
      The x-vector and RNN systems, as deep-learning-based models, need a large amount of training data to be able to generalize well.

    \item {\bf Explainability:} deep learning models are black-box models, being very difficult to interpret from a human perspective. On the contrary, the linguistic, fluency, functionals and i-vector  features created here are explainable and one could determine the weight of them in the classification.
\end{itemize}

Several future research lines have been identified for further work. Firstly, investigation on improved classification based on deep neural networks and novel acoustical feature  extraction algorithms. Secondly, addition of new linguistic features and non-verbal information (breaks, silence duration, word mistakes, etc.) in text-based systems. Thirdly, analysis of strategies for increasing the generalization capacity of the proposed speech-based systems; for example: to find new rhythmic parameters more discriminatory between patients with and without AD. It is also important to conduct experiments in other experimental settings, for example using other questions of the mini-mental state examination test, to validate the results obtained and, above all, to increase the size of the data settings.

\section{Acknowledgements}
This work has received financial support from the Spanish “Ministerio de Economia y Competitividad” through the project Speech\&Sign RTI2018-101372-B-100, and also from Xunta de Galicia (AtlanTTic and ED431B 2018/60 grants) and European Regional Development FundERDF.

\bibliographystyle{IEEEtran}

\bibliography{mybib}

\end{document}